\documentclass[10pt]{iopart}
\usepackage{psfig}  
\begin{document}

\def\NZ{{\bar N_0}}
\def\NJ{N_{\J}}
\def\J{J/\psi}
  
\title{Formation of quarkonium 
states at RHIC}

\author{R L Thews, M Schroedter and J Rafelski
}

\address{Department of Physics, University of Arizona,
Tucson, AZ 85721, USA}

\begin{abstract}
At RHIC the cross section for $c\bar{c}$ production 
will be large enough such that approximately 10 pairs will be
produced in each central collision.  
If a region of deconfined quarks and gluons is subsequently
formed, one would expect that the mobility of the charm
quarks will enable them to form $\J$ through ``off-diagonal"
combinations, involving 
a quark and an antiquark which
were originally produced in separate incoherent interactions.
We present model estimates of this effect, which indicate that
the signal for deconfinement at RHIC may possibly be $\J$ 
enhancement rather than suppression.
\end{abstract}

\section{Introduction}

A decrease in the number of observed $\J$ in heavy ion collisions 
due to the screening of the color confining potential was
proposed many years ago \cite{matsuisatz} as a signature of a deconfined 
phase.
It is argued that as the system cools and the deconfined phase disappears, these heavy
quarks will most likely form a final hadronic state with
one of the much more numerous light quarks.  The result will be a
decreased population of $\J$ relative to those formed initially
in the heavy ion collision.  

Here we study a scenario which can only be realized at RHIC (and LHC)
energies, where the average number of initially-produced 
heavy quark pairs $\NZ$ is 
substantially above unity in each central collision.  Then one can
amplify the probability of $\J$ formation by a factor which is
proportional to $\NZ^2$, {\it if and only if a space-time region
of deconfined quarks and gluons is present.}  Realization of this
result will depend on the efficiency of this new formation 
mechanism during the deconfinement period.  We have developed a
simple model to estimate the magnitude of this effect, and 
examined the sensitivity of results to various input parameters
and assumptions.

\section{Suppression Factor}

For expected conditions at RHIC, almost all of the directly-produced
$\J$ will be dissociated even in peripheral collisions.
To include the effects of our new
formation mechanism, we parameterize the final $\J$ number
in each event as follows:   

Of the $N_0$ 
charm quark pairs initially produced in a central heavy ion
collision, let $N_1$ be the number
of those pairs which form $\J$ states in the normal confining
vacuum potential.  At hadronization, the final number $\NJ$  
will contain a small fraction $\epsilon$ of the initial number $N_1$. 
The majority of $\NJ$ will 
be formed by this new mechanism which we expect to be
quadratic in the remaining 
$N_0 - N_1$ heavy quark pairs, with a proportionality parameter
$\beta$.  (We include in the new mechanism both
formation and suppression effects, since they occur simultaneously in
the deconfined region.)   
The final population is then
  
\begin{equation}
\NJ = \epsilon N_1 + \beta (N_0 - N_1)^2.
\label{eqquad}
\end{equation}
  
For each $N_0$ initially-produced heavy
quark pairs, we then average over the distribution of $N_1$, 
introducing the probability $x$ that a given heavy quark pair
was in a bound state before the deconfined phase 
was formed.  (This factor includes
the effect of interactions with target and projectile
nucleons).   
We finally average over the distribution of $N_0$, using a Poisson
distribution with average value $\NZ$, to obtain the
expected $<\NJ>$ final population per collision, 

\begin{equation}
<\NJ> = x \NZ (\epsilon + \beta(1-x)) + \NZ (\NZ+1)\beta (1-x)^2.
\label{eqave}
\end{equation}

The bound state ``suppression" factor $S_{\J}$ is just the ratio of
this average population to the average initially-produced
bound state population per collision, $x \NZ$.

\begin{equation}
S_{\J} = \epsilon + \beta (1-x) + \beta {(1-x)^2\over x} (\NZ +1)
\label{eqsupp}
\end{equation}

Without the new production mechanism, $\beta = 0$ and the 
suppression factor is $S_{\J} = \epsilon < 1$.
(Even the fitted parameter $\epsilon$
contains some effects of the new mechanism, since formation
can reoccur subsequent to the dissociation of an initial $\J$. Here we
use it as an upper limit with which to compare the complete result.)
However, it is 
possible that for sufficiently large values of $\beta$ 
and $\NZ$ this
factor could actually exceed unity, i.e. one would predict
an {\bf enhancement} in the heavy quarkonium production
rates to be the signature of deconfinement!
We thus proceed to estimate expected $\beta$-values
for $\J$ production at RHIC.

\section{Model for $\J$ Formation}

This model is adapted from our previous work on the formation
of $B_c$ mesons \cite{BC}.  Initial results for the $\J$ application
are found in Reference \cite{PRL}.
For simplicity, we assume the deconfined phase is an ideal gas of free
gluons and light quarks.  Any $\J$ in this medium will be subject
to dissociation via collisions with gluons.  (This is 
the dynamic counterpart of the plasma screening scenario, in which
the color-confinement force is screened away in the hot dense plasma
\cite{Kha}.) 
The primary formation mechanism is just the reverse of
the dissociation reaction, in which a free charm quark and antiquark 
are captured in the
$\J$ bound state, emitting a color octet gluon.  
Thus it is unavoidable for this model of quarkonium suppression that
a corresponding mechanism for quarkonium production must be present.
The competition
between the rates of these reactions integrated over the lifetime of
the QGP then determines the final $\J$ population.  Note that in
this scenario it is impossible to separate the formation process
from the dissociation (suppression) process.  Both processes occur
simultaneously, in contrast to the situation in which
the formation only occurs at the initial times before the QGP is
present.  

The time evolution if the $\J$ population
is then given by

\begin{equation}\label{eqkin}
\frac{d\NJ}{d\tau}=
  \lambda_{\mathrm{F}} N_c\, \rho_{\bar c } -
    \lambda_{\mathrm{D}} \NJ\, \rho_g\,,
\end{equation}                                                                                                                  
where $\tau$ is the proper time
in a comoving volume cell and  $\rho_i$ denotes the
number density $[L^{-3}]$ of species {\it i}.
The reactivity $\lambda$ $\left[L^3/\mathrm{time}\right]$ is
the reaction rate $\langle \sigma v_{\mathrm{rel}} \rangle$
averaged over the momentum distribution of the initial
participants, i.e. $c$ and $\bar c$ for $\lambda_F$ and
$\J$ and $g$ for $\lambda_D$.
The gluon density is determined by the equilibrium value in the 
QGP at each temperature.  
Exact charm conservation is enforced throughout
the calculation. 
The initial volume at
$\tau = \tau_0$ is
allowed to undergo longitudinal expansion 
$V(\tau) = V_0 \tau/\tau_0$.
The expansion is taken to be isentropic, $VT^3$ = constant,
which then provides a
generic temperature-time profile. For simplicity, we assume the 
transverse spatial distributions are uniform, and use a thermal 
equilibrium momentum 
distribution for both gluons
and charm quarks. (This last simplification requires large
energy loss mechanisms for the charm quarks in the deconfined
medium, which is indicated by several recent studies \cite{energyloss}).

 With these inputs and assumptions, the solution of Equation \ref{eqkin}
is precisely that anticipated in Equation \ref{eqquad}, with

\begin{equation}
\epsilon(\tau_f) = e^{-\int_{\tau_0}^{\tau_f}{\lambda_{\mathrm{D}}\, \rho_
g\,
d\tau}},
\label{eqalpha}
\end{equation}
where $\tau_f$ is the hadronization time determined by the
initial temperature ($T_0$ is a variable parameter) and
final temperature ($T_f$ = 150 MeV ends the deconfining phase),
and

\begin{equation}
\beta(\tau_f) = \epsilon(\tau_f) \times  \int_{\tau_0}^{\tau_f}
{\lambda_{\mathrm{F}}\, [V(\tau)\, \epsilon(\tau)]^{-1}\, d\tau}.
\label{eqbeta}
\end{equation}

For our quantitative estimates,
we utilize a cross section for the dissociation of $\J$
due to collisions with gluons
which is based on the operator product expansion
\cite{Kha,OPE},
which is utilized with detailed balance factors
to calculate the primary formation rate for the capture of 
a charm and anticharm quark into the $\J$.

\section{Results}

\begin{figure}[h]
\centerline{  \hspace*{-.cm}
\psfig{width=8.5cm,figure=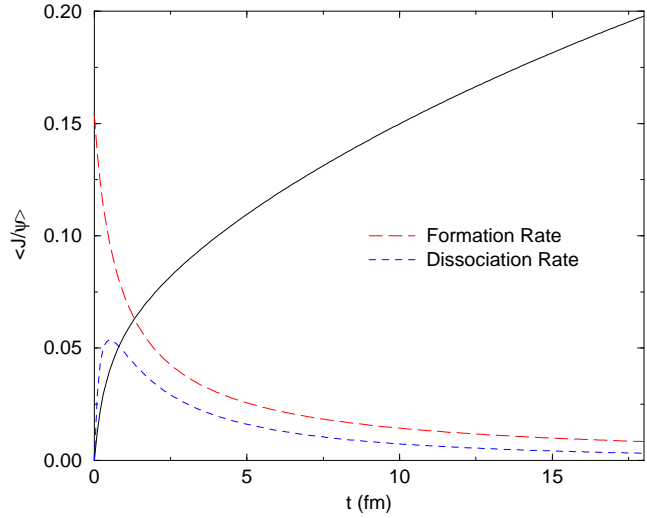}
}
\caption{ \small
Time dependence of $\J$ formation including new mechanism.
\label{jpsitime}}
\end{figure}

In Figure \ref{jpsitime} we show the time development of the
$\J$ (solid line) along with the separate formation and dissociation
rates (dotted lines, arbitrary units).  

This calculation maintained exact charm conservation, so that the
solutions followed evolution of both bound and free charm quarks.
One sees the expected
decrease of the formation rate due to the volume expansion, and
the decrease of the gluon dissociaton rate due to the decrease
in gluon density with temperature.

\begin{figure}[h]
\centerline{  \hspace*{-.cm}
\psfig{width=8.5cm,clip=,figure=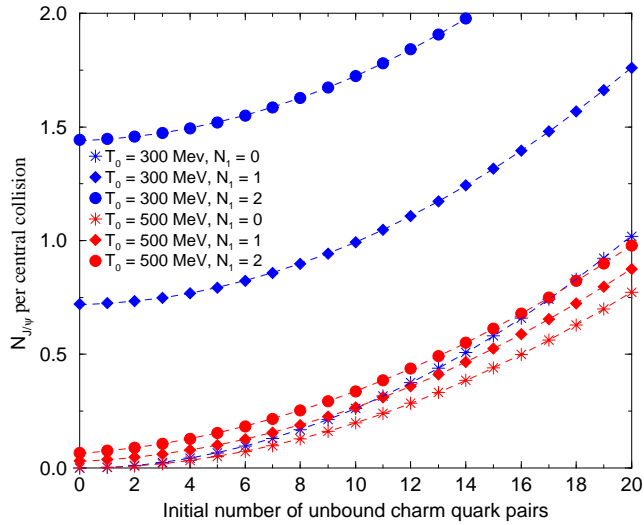}
}
\caption{ \small
Calculated $\J$ formation in deconfined matter  at several
initial temperatures at RHIC, as a function of initial charm production.
\label{figquad}}
\end{figure}


Some typical calculated values
of the $\J$ final population are shown in
Figure \ref{figquad}. The parameter values for thermalization
time $\tau_0$ = 0.5 fm, initial volume $V_0 = \pi R^2\tau_0$ with
R = 6 fm, and a range of initial temperature 300 MeV $< T_0 <$ 500 MeV, 
are all
compatible with expectations for a central collision at RHIC.
The quadratic fits of Equation \ref{eqquad}
are superimposed, verifying our expectations that the decrease in
initial unbound charm is a small effect.
(These fits also contain a small linear term
for the cases in which $N_1$ is nonzero, which accounts for
the increase of the unbound charm population when dissociation occurs.)
The fitted $\epsilon$ values decrease quite rapidly with increasing
$T_0$ as expected, and give reasonable upper limits 
for the 
suppression factor of directly-produced 
$\J$ in central collisions at RHIC due to
gluon dissociation in a deconfined phase.  The corresponding
$\beta$ values are relatively insensitive to $T_0$, remaining in the
range $2.0 - 2.6 \times 10^{-3}$.  These fitted parameters 
must be supplemented by values of $x$ and $\NZ$ to 
determine the ``suppression" factor from Equation \ref{eqsupp} for the
new mechanism.  We use $\NZ$ = 10  from a pQCD estimate 
\cite{hardprobes1}. An order of magnitude estimate of
$10^{-2}$ for $x$, from fitted values of a color evaporation
model \cite{hardprobes2}, is reduced by 
the
suppression due to interactions with target and beam nucleons.  For
central collisions we use 0.6 for this factor, which results
from the extrapolation of the observed nuclear effects for
p-A and smaller A-B central interactions.

 With these parameters fixed, we predict from Equation \ref{eqsupp}
an {\bf enhancement} factor
for $\J$ production of  
$3.6 < S_{\J} < 5.4$, for initial temperatures between
300 and 500 MeV.  
The suppression of initially-produced $\J$ alone ranges from 
factors of 10 to 100, so that the enhancement prediction involves 
a huge increase (factors of approximately one to two orders of
magnitude) in the final population of $\J$ at RHIC. 

\section{Centrality Dependence}
We also predict how this new effect will vary with the
centrality of the collision, which has been a key
feature of deconfinement signatures analyzed at
CERN SPS energies \cite{NA50}.  
The $\epsilon$ and $\beta$ parameters are recalculated, using
appropriate variation of initial conditions with impact parameter b.
From nuclear geometry and the total non-diffractive nucleon-nucleon
cross section at RHIC energies, one can estimate the total
number of participant nucleons $N_P(b)$ and the corresponding
density per unit transverse area $n_P(b,s)$ \cite{wounded}.
The former quantity has been shown to be directly
proportional to the total transverse energy produced in
a heavy ion collision \cite{ET}.  The latter quantity is used, along
with the Bjorken-model estimate of initial energy density
 \cite{Bj}, to provide an estimate of how the initial temperature
of the deconfined region varies with impact parameter.  We
also use the ratio of these quantities
to define an initial transverse area 
within which deconfinement is possible, thus completing the
initial conditions needed to calculate the $\J$ production
and suppression.  The average initial charm number
$\NZ$ varies with impact parameter in proportion to the
nuclear overlap integral $T_{AA}$(b).  The impact-parameter
dependence of the fraction $x$ is determined by the
average path length encountered by initial $\J$ as they
pass through the remaining nucleons, $L(b)$ \cite{gerschel}.
All of these b-dependent effects are normalized to the previous values 
used for calculations at $b = 0$.  

\begin{figure}[h]
\centerline{  \hspace*{-.cm}
\psfig{width=8.5cm,figure=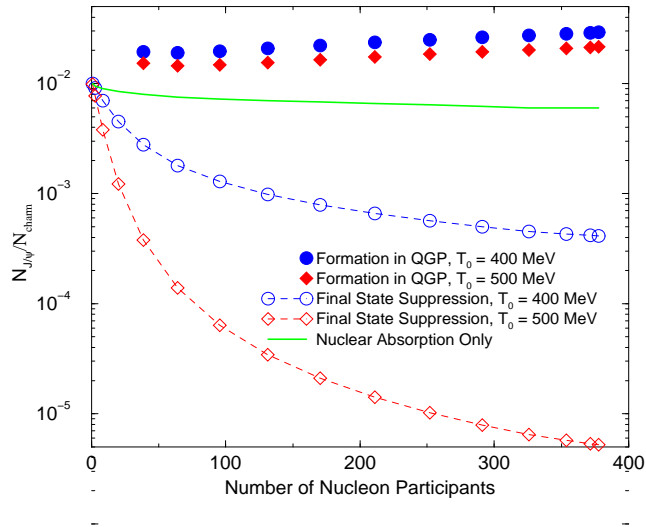}
}
\caption{ \small
 Ratio of final $\J$ to initial charm as a function of
centrality,
due to nuclear absorption only (solid line), after final state
suppression by a QGP (dashed lines), and with inclusion of the
new formation mechanism in the deconfined medium (solid symbols).
\label{figjpsib}}
\end{figure}
 
It is revealing to express these
results in terms of the ratio of final $\J$ to initially-produced
charm pairs.
In Figure \ref{figjpsib}, the solid symbols are the full results predicted 
 with the inclusion of 
our new production mechanism at RHIC. The centrality
dependence is represented by the total
participant number $N_P$(b). For comparison we also show
predictions without the new mechanism, when only dissociation by gluons is
included ($\lambda_F$ = 0).  It is evident not only that the
new mechanism dominates the $\J$ production in the deconfined medium
at all impact parameters, 
but also that 
an increase rather than
a decrease is predicted for central collisions.  These features should 
be distinguishable in the upcoming RHIC experiments.

\section{Model Dependence}

\begin{figure}[h]
\centerline{  \hspace*{-.cm}
\psfig{width=12cm,figure=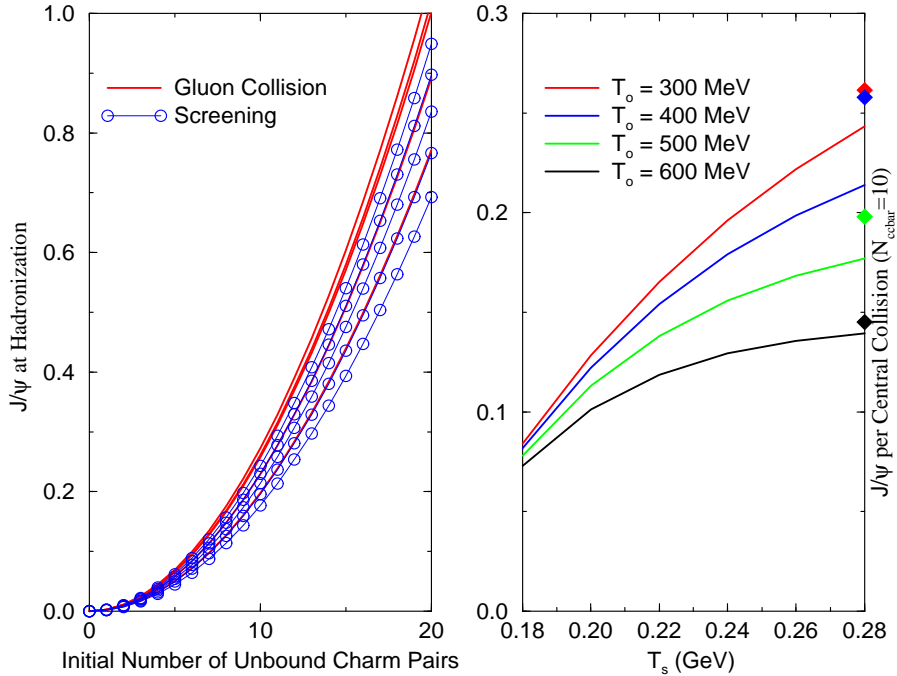}
}
\caption{ \small
Comparison of Gluon Dissociation and Screening Scenarios for
$\J$ formation including new mechanism (see text for details).
\label{jpsiscreen}}
\end{figure}

In our model of a deconfined region, we have used the vacuum 
values for masses and binding energy of $\J$, and assumed that the
effects of deconfinement are completely included by the dissociation
via gluon collisions.  For a complementary viewpoint, we have also
employed a deconfinement model in which the $\J$ is completely
dissociated when temperatures exceed some critical screening
value $T_s$.  Below that temperature, the new formation mechanism
will still be able to operate, and we use the same cross sections
and kinematics.  We find that for $T_s$ = 280 MeV, the final $\J$ population
is approximately unchanged, while decreasing $T_s$ to 180 MeV
could reduce the $\J$ production by factors of 2 or 3.
These results are shown in Figure \ref{jpsiscreen}.

We have also checked the sensitivity of these results to several
other assumptions and parameters.  Among these are: (a) Change in
initial charm production due to gluon shadowing; (b) Alternative
cross sections with different magnitudes and threshold behaviors;
(c) Transverse expansion of the QGP; (d) Non-chemical equilibrium
for gluons; (e)Non-thermal momentum distributions for charm
quarks.  The effects of varying these assumptions produce both
positive and negative changes in the final $\J$ populations.  The
largest effect could be a decrease by a factor of 2 or 3 if one
uses the initial pQCD momentum distributions for the charm quarks.
Taken together, however, it is unlikely that a conspiracy of
these effects would qualitatively change the predicted 
enhancement effects of this deconfinement scenario.

\section{Summary}

In summary, we predict that at RHIC energies the $\J$ production
rate will provide a more interesting signal for deconfinement than
has been previously realized.  Consideration of multiple heavy quark
production made possible by higher collision energy effectively
adds another dimension to the parameter space within which one
searches for patterns of quarkonium behavior in a 
deconfined medium. It will be possible to experimentally
``tune" the number of initial heavy quark pairs by sweeping through
either centrality or energy.  One can then search for a $\J$ production
behavior which is predicted to be nonlinear in total charm.
In our simplified kinetic model of $\J$ formation in a free
gas of quarks and gluons, the new production
mechanism predicts an enhancement rather than a 
suppression.
These features should provide a signal at RHIC which will be 
difficult to imitate with conventional hadronic
processes.  The extension of this scenario to LHC
energies will involve hundreds of initially-produced
charm quark pairs and multiple bottom pairs.
We expect the effects of this
new production mechanism to be striking.

{\ack Acknowledgment: This work was supported  by a grant from 
the U.S. Department of Energy,  DE-FG03-95ER40937.}



\section*{References}

\begin{thebibliography}{99}


\bibitem{matsuisatz} Matsui T and Satz H 1986  
{\it Phys. Lett.} {\bf B178} 416 

\bibitem{BC}
Schroedter M, Thews R L and Rafelski J 2000
{\it Phys. Rev.} {\bf C62} 024905 

\bibitem{PRL}
Thews R, Schroedter M and Rafelski J 2000
{\it Preprint} hep-ph/0007323

\bibitem{Kha}
Kharzeev D and Satz H 1994
{\it Phys. Lett.} {\bf B334} 155 

\bibitem{energyloss}
For a review, see Baier R, Schiff D and Zakharov B 2000
{\it Preprint} hep-ph/0002198

\bibitem{OPE}
Peskin M E 1979 {\it Nucl. Phys.} {\bf B156}, 365;
Bhanot G and Peskin M E 1979 {\it Nucl. Phys.}
 {\bf B156} 391

\bibitem{hardprobes1}  McGaughey P L, Quack E, Ruuskanen P V, 
Vogt R and Wang X -N 1995
           {\it Int. J. Mod. Phys.} {\bf A10} 2999 

\bibitem{hardprobes2} Gavai R, Kharzeev D, Satz H, Schuler G,
Sridhar K and 
Vogt R 1995
           {\it Int. J. Mod. Phys.} {\bf A10} 3043 

\bibitem{NA50} NA50 Collaboration, Abreu M C {\it et al.} 2000
{\it Phys. Lett.}
{\bf B477} 28 

\bibitem{wounded} Bia{\l}as A, Bleszy\'{n}ski M and Czyz W 1976
{\it Nucl.
Phys.} {\bf B111} 461 

\bibitem{ET}  Margetis S {\it et al.} 1995 {\it Nucl Phys.}
{\bf A590} 355c 

\bibitem{Bj}
Bjorken J D 1983
{\it Phys. Rev.} {\bf D27} 140 

\bibitem{gerschel}  Gerschel C and H\"{u}fner J 1992
{\ Z. Phys.} {\bf C47} 171 


\end{thebibliography}
\end{document}